\documentstyle[aas2pp4]{article}

\def\nhi{\mbox{$N_{\rm HI}$}}

\def\q0{q$_0$}
\def\lya{Ly$\alpha$\ }
\def\lyanv{Ly$\alpha+$NV}
\def\etal{\rm et al.}
\def\eg{\protect\rm e.g.}

\def\simlt{$_<\atop{^\sim}$}
\def\simgt{$_>\atop{^\sim}$}

\def\vs{\vspace{1ex}}
\def\notes{\vs\hspace{10mm}}
\def\hi{\mbox{\rm HI}}
\def\heii{\mbox{\rm HeII}}
\def\civ{\mbox{\rm CIV}}
\def\cii{\mbox{\rm CII}}
\def\oi{\mbox{\rm OI}}
\def\o1{\mbox{\rm OI}}
\def\si2{\mbox{\rm SiII}}
\def\soiv{\rm SiIV+OIV]}
\def\siv{\rm SiIV}
\def\os2{\rm OI+SiII}
\def\lyb{Ly$\beta$}
\def\nv{\rm NV}
\def\niv{\rm NIV]}
\def\ovi{\rm OVI}
\def\oiii{\rm OIII]}
\def\mgii{\rm MgII}
\def\feii{\rm FeII}

\def \m.        {\rlap{$.$}^{\rm m}}
\def \s.        {\rlap{$.$}^{\rm s}}
\def \am.       {\rlap{$.$}'}
\def \as.       {\rlap{$.$}''}

\slugcomment{in press, Astrophysical Journal Supplements, 1 September 1996}
\lefthead{Storrie-Lombardi, McMahon, Irwin, \& Hazard}
\righthead{APM z$\ge$4 QSO Survey: Spectra \& Intervening Absorption Systems}

\begin{document}

\title{\protect APM z \simgt\ 4 QSO Survey: \\ Spectra and Intervening Absorption Systems}

\author{L. J. Storrie-Lombardi\altaffilmark{1}, R. G. McMahon}
\authoraddr{UCSD-CASS, Mail Code 0111, 9500 Gilman Drive, La Jolla, 
CA  92093 USA}
\affil{Institute of Astronomy Madingley Road, Cambridge CB3 0HA, UK, \\
lsl@ucsd.edu and rgm@ast.cam.ac.uk}
\altaffiltext{1}{current address: UCSD-CASS, Mail Code 0111, 
9500 Gilman Drive, La Jolla, CA  92093 USA}
\authoremail{lsl@ucsd.edu}

\author{M. J. Irwin}
\affil{Royal Greenwich Observatory Madingley Rd, Cambridge CB3 0EZ, UK,
mike@ast.cam.ac.uk}

\and
\author{C. Hazard}
\affil{University of Pittsburgh, Pittsburgh, PA 15260 USA and 
Institute of Astronomy, hazard@ast.cam.ac.uk}

\author{arch-ive/9604021 | to appear in ApJ Supplements, 1 September 1996}

\begin{abstract}

The APM multicolor survey for bright z $>$ 4 objects, covering
2500 deg$^2$ of sky to m$_r \sim$ 19, resulted in the 
discovery of thirty-one quasars with z \simgt\ 4.  
High signal-to-noise
optical spectrophotometry at 5\AA\ resolution has been obtained for 
the twenty-eight quasars easily accessible from the northern hemisphere.
These spectra have been surveyed to create new samples of 
high redshift Lyman-limit systems, damped Lyman-$\alpha$
absorbers, and metal absorption systems (\eg\ \civ\ and \mgii).  
In this paper we present the spectra, together with line lists of the 
detected absorption systems.
The QSOs display a wide variety of emission and absorption
line characteristics, with 5 exhibiting broad absorption lines
and one with extremely strong emission lines (BR2248$-$1242). 
Eleven candidate damped Ly$\alpha$ absorption systems
have been identified covering
the redshift range $2.8\le z \le 4.4$ (8 with $z>3.5$).
An analysis of the measured redshifts of the high ionization 
emission lines with the low ionization lines shows them to be 
blueshifted by $430\pm60$ km s$^{-1}$.  In a previous
paper (\cite{SMIH94}) we discussed the redshift evolution 
of the Lyman limit systems catalogued here. In subsequent papers 
we will discuss the properties of the \lya forest absorbers 
and the redshift and column density evolution 
of the damped \lya absorbers.

\end{abstract}

\date { today }


\section{Introduction}

Many QSOs have now been discovered beyond redshifts of 4
and they provide powerful probes for exploring these early epochs.
They are the youngest objects known in the Universe and
it is likely that they flag regions where
galaxy formation is very active.  Their host galaxies are probably
still forming and they may occur in the exceptional
`5 $\sigma$' peaks in the matter distribution of the early
Universe (Efstathiou \& Rees 1988).
Observational information from this epoch yields 
constraints on galaxy formation theories and clues
for better understanding of the
astrophysics of galaxy formation and evolution.

In addition to being of intrinsic interest themselves,
bright high redshift QSOs are particularly valuable as probes of
the intervening gas clouds and galaxies superimposed on
their spectra in absorption.
The study of these absorption systems provides
information about the formation and evolution of galaxies over
most of the age of the Universe.
Neutral hydrogen (\hi) absorption
can be detected over a staggering 10 orders of magnitude
from the \lya forest region with the weakest detectable lines
having a column density \nhi\ $\sim$ $10^{12}$ atoms cm$^{-2}$, 
up to the damped \lya absorbers
with \nhi\ $\sim$ $10^{21}$.  
The rich zoo of these absorbers, in addition to those produced
by heavier elements such as carbon, silicon, oxygen, and magnesium,
are illuminated along a QSO line-of-sight, leaving their imprint
as absorption in the QSO continuum.
Study of the \lya forest (12 \simlt\ $\log$ \nhi\ \simlt\ 17)
yields important information
about the intergalactic medium and the
background ionizing flux at high redshifts (\eg\ Hunstead \etal\ 1986;
Carswell \etal\ 1987; Bajtlik, Duncan, \& Ostriker 1988; Williger \etal\ 1994).
Lyman-limit systems ($\log$ \nhi\ \simgt\ 17) provide a means of
directly studying the evolution of galaxies
over the redshift range 0.1 ${<}$ z ${<}$ 5 
(\eg\ Sargent, Steidel, \& Boksenberg 1989; 
Lanzetta \etal\ 1991; Storrie-Lombardi \etal\ 1994;
Stengler-Larrea \etal\ 1995).
The absorbers detected via the damped \lya lines they produce
($\log$ \nhi\ \simgt\ 20) show
features consistent with an early phase of galactic evolution
and are widely believed to be the progenitors of
spiral galaxies like our own (\eg\ Wolfe \etal\ 1986; Wolfe 1987;
Fall, Pei, \& McMahon 1989;
Pettini, Boksenberg, \& Hunstead 1990; Rauch \etal\ 1990;
Lanzetta \etal\ 1991; Lanzetta, Wolfe, \& Turnshek 1995;
Wolfe \etal\ 1995).

The APM Color Survey for z$>$4 QSOs was undertaken to find bright
(m$_{\rm R}$ \simlt\ $19$) quasars with redshifts 4 \simlt\ z \simlt\ 5 
(\cite{IMH91}).
The aim of the program was to find a large sample
of QSOs for both intrinsic and absorption line studies.
The survey covers 
approximately 2500 square degrees of sky from the equatorial region
of the UK Schmidt Telescope (UKST) B$_{\rm J}$, R, I Survey
with $\vert$b$\vert > 30^{\circ}$ and declination range $+3$ to $-17.5$.
High signal-to-noise optical spectrophotometry at 5\AA\ resolution covering
the wavelength region 3500$-$9000\AA\ were obtained 
for all the QSOs discovered in the APM Color Survey  
that are accessible using the William Herschel Telescope (WHT)
(28 of 31 objects).
In addition, spectra were obtained at 5\AA\ resolution 
of three high redshift radio-selected QSOs  
(Hook \etal\ 1995; McMahon \etal\, in preparation).
The spectra have been utilized to discover
Lyman-limit, damped Ly$\alpha$, and metal absorption
systems (\eg\ \civ\  and \mgii).  

The results and analyses from these studies are presented in a 
series of papers. In this paper we present the spectra, together with 
a list of accurate redshift determinations of the intrinsic QSO emission 
lines, a study of the velocity differences between high and low
ionization emission lines, and the results of surveys for 
intervening absorption systems. Line lists for damped \lya candidates, 
Lyman limit systems, and metal absorption systems  
are provided.  The analysis of the redshift evolution 
of the Lyman limit systems was previously
presented in Storrie-Lombardi \etal\ (1994). 
We will present a detailed analysis of 
the damped \lya systems and the redshift evolution 
of their number density and column density distribution (\cite{SIM95}),
and describe the implications derived from the 
damped \lya survey for the evolution of  
the mass density of neutral gas with redshift and 
the implications for galaxy formation (\cite{SMI95}).
Other papers will cover studies of the \lya forest clouds at high redshift
and the intrinsic properties of the QSOs.   
High resolution studies of the \lya forest region at $z>4$ have
been completed by Williger \etal\ (1994) and Wampler \etal\ (1996).

\section{Observations}

High signal-to-noise optical spectrophotometry at 5\AA\ 
resolution covering
the wavelength range 3500$-$8800\AA\ was obtained
with the 4.2m William Herschel telescope
of the Isaac Newton Group of telescopes in the Canary Islands.
We used the ISIS double-spectrograph with typical integration
times of 2700-3600 seconds.
The spectrophotometry is accurate to within 5-10 percent.
ISIS is a double beam spectrograph with arms optimized for blue and red
light, mounted at the f/11 Cassegrain focus of the WHT. 
For this project the lowest dispersion was required, and 
gratings with 158 lines/mm and a dichroic to split the light at
$\sim$5400\AA\ were used. This gives 2.71\AA/pixel in the red arm
and 2.89\AA/pixel in the blue. The gratings were arranged so that the blue
part of the spectrum was centered on 4600\AA\ and covered a range of
2950\AA\ while the red was centered on 7000\AA\ and covered a range of
3380\AA.  The red and blue arm observations were carried out
simultaneously.
On the red arm an English Electric Valve (EEV)
$1242\times 1152$ CCD with $\rm 22.5\mu m$ pixels
was used as detector. On the
blue arm a thinned Tektronix $1024\times 1024$
CCD with $\rm 24\mu m$ pixels was used.
All the narrow slit observations were taken with a slit width 
of 1.5" except
for BR $2237-0607$ (1") and all were taken with the slit perpendicular
to the horizon (at the parallactic angle). This slit orientation is used
to minimize the effects of atmospheric differential refraction
(\cite{Filippenko82}). 

Though the QSOs from the APM Color Survey are bright
for high redshift objects (m$_{\rm R} < 19$), they were
barely visible on the acquisition TV, so nearby offset stars
(m$_{\rm R} \sim 15-17$) were acquired first.  Blind-offsetting
was then used to position the target object in the slit and as a check
the TV integration time was increased until the periphery of the 
QSO was visible in the slit.
All observations were made with a long slit and the CCDs were
windowed in the spatial direction to reduce the overhead 
due to readout time.
The observations are summarized in table 1.

\section{Data Reduction}

The data were reduced using
standard software from the IRAF\footnote{IRAF is distributed by the
National Optical Astronomy Observatories, which is operated by the
Association of Universities for Research in Astronomy, Inc.~(AURA)
under cooperative agreement with the National Science Foundation.}
package. After the data were overscan, bias, and flat-field corrected   
they were extracted using the variance-weighted extraction in APALL. 
Typically the ISIS spectra curved by less than a pixel from one 
end of the chip to the other.  APALL outputs the sky spectrum which 
was used for quick wavelength calibration at the telescope.  
A sample dark sky spectrum taken with ISIS is shown
in figure~\ref{f_sky}. 
CuNe$+$CuAr arcs were taken at intervals throughout 
the night to provide an accurate wavelength calibration. 
Emission lines were identified and a pixel-to-wavelength calibration
curve was found by fitting a 3rd order chebyshev polynomial to
the calibration points in IDENTIFY. Typical {\it rms} residuals
from the fit were 0.2\AA.
The dispersion solution was applied to the extracted spectra and they
were put on a linear wavelength scale using DISPCOR.

The individual spectra were then extinction corrected
and coadded.
Spectrophotometric standards taken from Oke (1974) and 
Oke \& Gunn (1983)  
were used to flux calibrate the spectra.
The goal was absolute spectrophotometry
correct to within 10 percent and relative flux levels from
the blue and red arms accurate enough to allow determination of
the spectral indices.
The flux calibration procedure was checked by flux-calibrating the
standards and overlaying calibration points.  It was found that
calibration was reliable over the wavelength range 5500\AA-8600\AA\
(ISIS red) and 3500\AA-5500\AA\ (ISIS blue).
Observations with a 5" slit were obtained for all but three 
of the QSOs. These were reduced in the same way as the narrow 
slit observations and used to correct the absolute flux 
levels for slit losses.  The slit losses ranged from 0$-$50\%.

`Featureless' B-stars were selected from the {\it Bright
Star Catalogue} (\cite{BrightStarCat}) or {\it Sky Catalogue 2000.0}
(\cite{Skycat}) for use in removing the effects of atmospheric 
absorption in the red spectra, (\eg\  O$_2$ A-band at 7600\AA).
Observations of B-stars were taken at different air masses
to provide a range of absorption so that as many objects as possible
could be corrected.
The spectrum of HR4468 taken at an airmass of 1.48 is shown in
figure~\ref{f_bstar2} and HD13679 taken at an
airmass of 1.06 is shown in figure~\ref{f_bstar}(a).
The atmospheric absorption features seen in the B-star spectrum were
removed by interpolating
between values on either side of the feature,
resulting in the spectrum shown in figure~\ref{f_bstar}(b).
The original
B-star spectrum was then divided by this featureless spectrum with
the result shown in figure~\ref{f_bstar}(c).
The object spectra taken at comparable air masses
were then divided by the result.
The technique was successful in almost
all cases though it remains an art form to do it properly.

The red and blue arm spectra were joined using SCOMBINE. 
The final reduced spectra are shown in
figure~\ref{f_data} and the individual QSOs
are described in section 5. 
The AB magnitude\footnote{
$AB = -2.5* \log (f(\nu)[ergs/s/cm^2/Hz)] - 48.6$ as defined
by Oke (1969).}
measured at
$\lambda_{rest}=1450$\AA\ and 
$\lambda_{observed}=7000$\AA,
along with the APM $R$ magnitudes measured from the plate scans
are listed in table 2. 
Using the absolute flux calibration for Vega taken 
from Hayes \& Latham (1975)
and defining the zero-point of the AB magnitude system 
at $\lambda = 5556$\AA\ 
leads to the following magnitude zero-point differences: AB measures at
$\lambda = 7000$\AA\ are 0.25 magnitudes fainter than on a Vega-based system;
for $\lambda = 1450\times(1+z)$\AA\ the difference ranges between 0.3 and
0.5 magnitudes for a 4 \simlt\ z \simlt\ 5 sample; the effective wavelength
of the photographic R-band is $\lambda = 6500$\AA\ and the difference between
the systems here is 0.2 magnitudes.  
The 1$\sigma$ errors are $\pm0.1$.
For those objects with no long slit observations 
or non-photometric conditions, only the APM R magnitude is quoted.  

\section{Redshift Measurements}

\subsection{Measuring the Emission Line Redshifts}

At redshifts greater than 4, Ly$\alpha+$\nv\ (rest wavelengths 1215.67\AA\
and 1240.13\AA) and \civ\ (1549.1\AA) are usually the only 
strong emission lines visible.
\lya is almost 50\% absorbed by the \lya forest, making it difficult
to use for redshift determination. Emission lines from
\lyb\ (1025.72\AA), \ovi\ (1034.0\AA), \si2\ (1263.0\AA),
\o1\ (1304.46\AA), \cii\ (1335.0\AA), \soiv\ (1400.0\AA), \niv\ (1486.0\AA),
\heii\ (1640.4\AA), and \oiii\ (1663.0\AA) may also be detected.
Single Gaussians were fit to the emission lines in each QSO, and
the redshifts for each line were determined from the central
wavelength ($z=\lambda_{observed}/\lambda_{emitted} - 1$).
For the Ly$\alpha+$\nv\ blend the fit was mainly to the red wing due
to the absorption by the forest.  The
redshifts for the strong metal lines were then averaged together 
(excluding Ly$\alpha$) to
determine a mean redshift for the QSO.
The redshift of each of the emission lines, the mean QSO redshift,
the 1$\sigma$ error, and the lines used in the determination
are shown in table 3.

\subsection{Emission Line Velocity Shifts}

There are uncertainties in the systemic redshift of the QSOs
in that redshifts determined
from high and low ionization lines have been shown to
exhibit velocity differences up to 2000 km s$^{-1}$ 
(\eg\ Espey \etal\ 1989; Steidel \& Sargent 1991; 
Carswell \etal\ 1991; Tytler \& Fan 1992),
where the velocity difference, $\Delta v$, is defined as 
$$ \Delta v = c {z_{ion} - z_{civ} \over (1 + z_{civ})}. $$
The same trend is exhibited in the APM sample, with a median difference of
$430\pm60$ km s$^{-1}$ between \oi\ and \civ, with the high
ionization line \civ\ blue-shifted with respect to the \oi.
The velocity difference relative to \civ\ has been calculated
for all the measured emission lines and the results are summarized
in table 4.
The BAL QSOs have been excluded from this analysis.
Histograms of the velocity differences are shown in
figure~\ref{f_vdiff}.  Some of the very large differences of several
thousand kilometers per second are due to the difficulty in
accurately measuring some of the heavily absorbed emission lines,
\eg\ Ly$\alpha$.
These shifts are important in estimating the metagalactic
ultraviolet background flux based on the proximity effect
(e.g. Williger \etal\ 1994) since an error in the redshift of the 
QSO of $\sim$1000 km s$^{-1}$ can lead to a factor of $2-3$ error in 
the derived ionizing flux. 

\section{Emission and Absorption Features}

The character of the emission lines and the \hi\ and metal 
absorption systems detected are described below for each QSO.  
Additional analysis of the intrinsic properties of the 
QSOs are described in a separate paper. 
The metal absorption systems in the non-BAL QSOs 
were selected with an automated algorithm that 
detected absorption features redward of \lya emission
with an equivalent width W $\ge$ 3$\sigma$ in 2.5 resolution elements.  
In the cases where the feature detected included more than
one line, Gaussians were fit to the lines individually to
measure the redshifts and equivalent widths. 
The results, with $1\sigma$ errors, are listed in table 5,
along with the identification of ion and redshift where possible. 
The selection of the damped \lya candidates is discussed in section 6 
and the Lyman limit systems previously published in Storrie-Lombardi
\etal\ (1994) are summarized in table 6.

\notes a) BR0019$-$1522, ($z_{em} = 4.528$)

The \lyanv\ and \civ\ emission lines are strong and sharp.  
There is a Lyman-limit system at z$=$4.27 and a damped \lya candidate
system at z$=$3.98. \si2, \civ, and \feii\ absorption are observed 
at z$=$3.4.  

\notes b) BRI0103$+$0032, ($z_{em} = 4.437$)

Strong and sharp \lya and \civ, with weaker \o1, \cii, and \soiv\
can be seen.  There is
weak \mgii\ at z$=$1.818 with 2 corresponding \feii\ lines
and \mgii\ at z$=$1.366. 
There are absorption edges visible just 
shortward of the emission lines that correspond to 
\os2 at z$=$4.41 and \siv\ and \civ\ at z$=$4.37.
There are two Lyman-limit systems at z$=$4.31 and 4.15. 

\notes c) BRI0151$-$0025, ($z_{em} = 4.194$)

The \lya and \civ\ emission lines are strong and sharp
with weaker \o1 and \soiv\ emission. 
There is a Lyman-limit system 
at z$=$4.05 and \civ\ at z$=$3.876.  At z$=$4.17
there is a strong \lya absorption line, a \nv\ doublet,
and a single line that could be \civ.
There is \mgii\ with at least one \feii\ line at z$=$1.91.

\notes d) BRI0241$-$0146, ($z_{em} = 4.053$)

The \lyanv, \o1, \cii, \soiv, and \civ\ emission lines are broad and 
rounded.  There is a strong Lyman-limit system at z$=$4.10.
There are numerous absorption lines redward of the \lya emission but
they are not easily identifiable. \mgii\ and \feii\ are 
identified at z$=$1.435.  Shallow absorption troughs 
are also visible.  

\notes e) BR0245$-$0608, ($z_{em} = 4.238$)

The \lyanv, \o1, \soiv, and \civ\ emission lines are weak. There
is strong \lya absorption on the blue side of the \lya emission  
corresponding to the Lyman-limit system at z$=$4.23. An \mgii\ doublet 
with 4 corresponding \feii\ lines is observed at z$=$1.711. 
There are strong, narrow \lya absorption lines (though not damped 
candidates) at z$=$3.36 and 4.14, with corresponding
\siv\ and \civ\ at z$=$4.14.  \civ\ is also detected at z$=$3.184.

\notes f) BR0351$-$1034, ($z_{em} = 4.351$)

This is one of the most unusual objects in the survey with 
saturated \civ\ absorption in the middle of the \civ\ emission. 
There are a large number of absorption lines including 
\siv\ at z$=$4.098 and 4.352, \civ\ at z$=$3.633, 4.098, 4.351, and 4.351,
\mgii\ at z$=$1.340 and 1.931, and \nv\ at z$=$4.353.
Due to the difficulty in measuring redshifts from the 
heavily absorbed emission lines, the redshift for this object was
calculated from the \civ\ absorption at z$=$4.351.

\notes g) BR0401$-$1711, ($z_{em} = 4.236$)

The \lyanv, \o1, and \civ\ emission lines are strong and sharp,
while the \soiv\ is broader and weaker.  The \o1\ is unusually
prominent.  There is strong absorption in the \civ\ emission
line at z$=$4.229. There is Lyman limit system at the QSO redshift.
The absorption feature at $\sim$7600\AA\ is
a residual from the removal of the atmospheric absorption line. 
The spectrum is very noisy at the red end of the blue arm 
portion ($\sim$5600\AA) and does not join together smoothly with
the red arm spectrum. 

\notes h) BR0945$-$0411, ($z_{em} = 4.145$, BAL)

This is the first of the five QSOs in the APM Color Survey 
that exhibits broad absorption lines.
\ovi, \nv, \siv, and \civ\ are observed at z$\approx$4.01.

\notes i) BR0951$-$0450, ($z_{em} = 4.369$)

The \lyanv, \o1, \cii, \soiv, and \civ\ emission lines are weak. There are 
damped \lya candidates at z$=$3.84 and 4.20. \civ\ doublets are 
identified at z$=$3.703, 3.855, 4.196, and 4.364 and \siv\ at z$=$3.703 and
3.858. There is a Lyman-limit system at z$=$4.22. 

\notes j) BRI0952$-$0115, ($z_{em} = 4.426$)

This QSO is gravitationally lensed (\cite{Lens0952}).  The \lyanv, \soiv, 
and \civ\ emission are weak and heavily absorbed.  There is a strong
damped \lya candidate systems at z$=$4.01. 
\civ\ doublets are identified at z$=$3.294, 3.475, 3.719 and 4.023
and \mgii\ at z$=$1.993. There is a Lyman-limit system at z$=$4.25.

\notes k) BRI1013$+$0035, ($z_{em} = 4.405$)

The \lyanv, \o1, \cii, \soiv, and \civ\ emission lines are weak. There is a 
damped \lya candidate at z$=$3.10 with corresponding \feii\ detected. 
There is an \mgii\ doublet at z$=$2.054 with 6 corresponding
\feii\ lines at z$=$2.058. There is a Lyman-limit system at z$=$3.78.  

\notes l) BR1033$-$0327, ($z_{em} = 4.509$)

The \lyanv, \o1, \cii, \soiv, and \civ\ emission lines fall at the 
strong end of the weaker-lined objects.  
There is a Lyman-limit system at z$=$4.19 and \cii\ tentatively 
identified at z$=$4.148.
See Williger \etal\ (1994) for a detailed analysis of the \lya forest 
region in this object.

\notes m) BRI1050$-$0000, ($z_{em} = 4.286$)

The \lya and \civ\ emission lines are strong and sharp
with weaker \ovi, \o1, and \soiv\ emission. 
There is \cii, \siv, and \civ\ detected at z$=$3.862 and
a Lyman-limit system at z$=$4.08.

\notes n) BRI1108$-$0747, ($z_{em} = 3.922$)

The \lya and \civ\ emission are strong and this is one of the few objects
where \lyb, \ovi, and \nv\ are easily distinguished, along with \o1, 
\cii, \soiv, and \niv. 
\civ\ doublets are observed at z$=$3.575, and 3.607.  

\notes o) BRI1110$+$0106, ($z_{em} = 3.918$)

The \lyanv, \o1, \cii, \soiv, and \civ\ emission are weak.  The 
O$_2$ A-band atmospheric absorption has been removed from the 
\civ\ emission line and a residual spike was cut-off, resulting in 
the unreal flat top to the emission line.  \mgii\ and 2 \feii\
lines are observed at z$=$1.479 and  \mgii\ at z$=$1.800.  

\notes p) BRI1114$-$0822, ($z_{em} = 4.495$)

The \lyanv, \soiv, and \civ\ emission lines are weak though the 
\lyb $+$\ovi\ is fairly prominent.  There is a damped \lya candidate
at z$=$4.25 and a Lyman-limit system at z$=$4.51.
There is absorption in the blue wing of the \lya
emission line that corresponds to \mgii\ at z$=$1.395 but no
confirming \feii\ can be observed due to the \lya forest.
Single absorption lines are observed that could correspond to \siv\ and \civ\ 
at z$=$3.91 and \civ\ at 4.25.  \civ\ doublets are seen at z$=$3.422, 3.571,
and 3.589. There is an \mgii\ doublet at z$=$1.794.

\notes q) BR1117$-$1329, ($z_{em} = 3.958$, BAL)

This QSO exhibits broad absorption lines 
for \ovi, \nv, \siv, and \civ\ at z$=$3.62 and 3.89.

\notes r) BR1144$-$0723, ($z_{em} = 4.147$, BAL)

This object exhibits broad absorption lines but also has 
detectable intervening absorption. 
There is a strong broad absorption trough
corresponding to \civ\ and a weak trough for  
\siv\ at z$=$4.00.  The emission lines are weak.  
There is a damped \lya candidate system at z$=$3.26 
but this is probably confused with broad \ovi\ absorption at z$=$4.0.
There is \mgii\ absorption at z$=$1.905 and 5 corresponding \feii\ lines. 

\notes s) BR1202$-$0725, ($z_{em} = 4.694$)

This is the highest redshift and brightest object in the APM sample. 
It has very weak emission lines with \lya and \civ\ almost completely 
absorbed away.  The spectrum is very similar to that of BRI1335$-$0417, 
described below.  The redshift is determined from the edge of the 
\lya emission line since the metal lines are so heavily absorbed. 
(The redshift determined from \o1\ and \civ\ is 4.679.)
There is a damped \lya system at z$=$4.38
and a Lyman-limit
system at z$=$4.52. \mgii\ doublets, some with associated \feii, are observed 
at z$=$1.463, 1.754, 2.238, 2.339, and 2.444.
\civ\ is detected at z$=$3.525, 3.565, 4.474, and 4.679.
See Wampler \etal\ (1996) for a detailed analysis of 
the \lya forest in this object.
 
\notes t) BR1302$-$1404, ($z_{em} = 3.996$, BAL)

This QSO exhibits a complex series of broad absorption lines for 
\ovi, \nv, \siv, and \civ\ at z$\approx$ 3.65, 3.72, and 3.92.
There are two \mgii\ doublets at z$=$2.044 and 2.058 with 4 and 3 
associated \feii\ lines, respectively. 

\notes u) BRI1328$-$0433, ($z_{em} = 4.217$)

The \ovi, \lya, \nv, \soiv, and \civ\ emission lines are 
strong with weaker 
\o1 present.  This is one of the few objects in the sample with 
well defined \nv\ emission. There is strong \mgii\ absorption at z$=$1.628
with 2 \feii\ lines. 
Lyman-limit systems are seen at z$=$3.31 and 4.25.

\notes v) BRI1335$-$0417, ($z_{em} = 4.396$)

The \lyanv\ and \civ\ emission lines are very weak, with the \lya almost
completely absorbed away.  This QSO is looks very similar to 
BR1202$-$0725.  
\mgii\ with 4 associated \feii\ lines is seen at z$=$1.822.
There is a Lyman-limit system at z$=$4.45 and \si2\ and \cii\ at z$=$4.40.  
There is strong \lya absorption at the QSO redshift.

\notes w) BRI1346$-$0322, ($z_{em} = 3.992$)

The \lya and \civ\ emission lines are very strong and sharp. 
There is \nv\ absorption at z$=$3.974. 
There are \civ\ doublets at z$=$3.359 and 3.994, and a single 
line that is most likely \civ\ at z$=$3.974.
\mgii\ with \feii\ is seen at z$=$1.944.
There is a damped \lya candidate at z$=$3.73
and a corresponding Lyman-limit absorption edge at z$=$3.75.  

\notes x) BRI1500$+$0824, ($z_{em} = 3.943$)

The \lyanv, \o1, \soiv, and \civ\ emission lines are weak but sharp.
There is a damped \lya candidate at z$=$2.80, \mgii\ with 6 \feii\ lines
at z$=$1.908, and \civ\ absorption in the emission 
line at z$=$3.940. This object
shows one of the most successful removals of the O$_2$ A-band absorption
at 7600\AA, in the middle of the \civ\ emission. 

\notes y) GB1508$+$5714, $z_{em} = 4.283$

This is a radio-selected object from Hook \etal\ (1995).
The \lya and \civ\ emission lines are strong and sharp.
There is a Lyman-limit system at z$=$3.9.  

\notes z) MG1557$+$0313, $z_{em} = 3.891$ 

This is a radio-selected object from McMahon \etal\ (in preparation).
The \lya and \civ\ emission are strong and sharp.
There is \civ\ absorption at z$=$3.898.

\notes aa) GB1745$+$6227, $z_{em} = 3.901$

This is a radio-selected object from Hook \etal\ (1995), also
discovered independently by Becker, Helfand, \& White (1992) 
on the basis of its X-ray emission. 
The \lya and \civ\ emission are strong and sharp. 
It has \mgii\ absorption at z$=$1.471.  There are 6 \feii\ lines at z$=$2.322,
but the corresponding \mgii\ doublet is not seen as it should occur at
9296\AA, redward of the end of this spectrum.  

\notes bb) BR2212$-$1626, ($z_{em} = 3.990$)

The \lyanv\ and \civ\ emission lines are strong and sharp with
weaker \o1, \soiv, and \niv. 

\notes cc) BRI2235$-$0301, ($z_{em} = 4.249$, BAL)

This QSO is the highest redshift BAL in the sample and has 
very broad absorption troughs. The emission lines
are almost completely absorbed, making it difficult to determine an
accurate redshift. It exhibits broad
absorption lines of \ovi\ (z$=4.08$), \nv\ (z$=$3.74), 
\siv\ (z$=$3.83), and \civ\ (z$=$3.65, 3.82, 4.03). 
There is a possible \mgii\ doublet at z$=$1.873. 

\notes dd) BR2237$-$0607,  ($z_{em} = 4.558$)

The \lyanv, \o1, \soiv, and \civ\ emission are strong with the \lya line
being particularly sharp.  There is a damped \lya candidate at 
z$=$4.08 and 
a Lyman-limit system at z$=$4.28. There is a 
\nv\ doublet at z$=$4.545, \siv\ at z$=$4.079, \cii\ at z$=$4.078,
\civ\ at z$=$4.482, \feii\ at z$=$2.155, and \mgii\ at z$=$1.672. 

\notes ee) BR2248$-$1242, ($z_{em} = 4.161$) 

This QSO has pathologically
strong emission lines for \ovi+\lyb, Ly$\alpha$, \nv, \o1, \cii, \soiv, 
\niv, \civ, \heii, and \oiii. It is the only object with obvious \niv.  

\section{Survey for Damped Lyman-{\bf $\alpha$} Absorption Systems}

\subsection{Background}

While the baryonic content of spiral galaxies 
that are observed in the present epoch is concentrated in stars, 
in the past this must have been in the form of gas. The principal gaseous
component in spirals is neutral hydrogen which has led
to surveys for absorption systems detected by the damped \lya (DLA)
lines they produce (Wolfe \etal\ 1986; Lanzetta \etal\ 1991;
Lanzetta \etal\ 1995; Wolfe \etal\ 1995). 
Damped \lya absorption systems comprise the 
high column density tail of neutral hydrogen absorbers  
with column densities of  
\nhi\ $\ge 2 \times 10^{20}$ cm$^{-2}$.  
They are identified by the presence of broad
(FWHM$>$5\AA) absorption lines shortward of Lyman-$\alpha$ (1216\AA) in
the QSO rest frame. These lines are broadened by radiation damping
and at z$>$4 have observed equivalent widths of W \simgt\ 25\AA.
The visibility of the damping wings in the absorption profile is
due to the large \hi\ column density and the low velocity dispersion 
($\sim$10 km s$^{-1}$), two features that damped systems have
in common with spiral galaxies observed at the present.
The column density along a typical line-of-sight in the Milky Way is 
\nhi\ $\sim10^{21}$ atoms cm$^{-2}$.
Other features that resemble \hi\ disk galaxies are the 
presence of metals in mainly low ionization states such as 
C$+$, Si$+$, and Fe$+$ and the detection of 21 cm absorption 
associated with damped systems shows that the gas is cold and has
a low level of turbulence ({\it c.f.} Wolfe \etal\ 1986; 
Wolfe 1987; Turnshek \etal\ 1989).

\subsection{Selection of Damped \lya Candidates} 

The \lya forest region in QSO spectra at redshift 4 is
very crowded.  Many lines are blended at 5\AA\ resolution and may appear 
broader than they actually are. 
However, real damped absorbers at high redshift result in 
very broad lines. They have observed equivalent widths 
W $>$ 25\AA\ and are relatively easy to see in the spectra.  
Two techniques were used to select
the candidate absorbers. We first selected the candidates 
interactively and then independently used the standard equivalent
width selection criteria with an automated algorithm. 
There was good agreement between the two selection methods
though we only report the candidates selected with the automated 
algorithm.  This is described below.
  
The technique used for selecting candidates and measuring
the sensitivity of the survey with redshift
follows the methods described in Lanzetta \etal\ (1991) 
and is described below. A local continuum was fit to each
spectrum using straight lines between the peaks of the forest
regions.  
An equivalent width spectrum and variance spectrum were created 
for each QSO defined as 
$$ W_i = \Delta \lambda \sum^{i+m}_{n=i-m} (1 - F_n/C_n), {\rm and}$$
$$ \sigma^2_{E_i} = (\Delta \lambda)^2 \sum^{i+m}_{n=i-m} (\sigma_{F_n}/C_n)^2,$$
where $C_i$ and $F_i$ are the continuum and flux levels at pixel $i$, 
$\Delta \lambda$ is the \AA$/$pixel of the spectrum, and $2m+1$ is the passband over
which the total equivalent width is measured. A passband of 15 pixels 
was used, equivalent to 37.5\AA. 

The spectra were analyzed using the above algorithm 
starting 3000 km s$^{-1}$ blueward of the emission redshift to
avoid lines possibly associated with the quasar.    
The analysis was stopped when the signal-to-noise ratio became 
too low to detect a \lya line with W(rest)$\ge$5\AA\ at 
the $5\sigma$ level.  
This point was typically caused by the incidence of a Lyman limit system.
This selected wavelength range is used to construct the   
sensitivity function, $g(z)$.  It gives the number of lines of sight at 
a given redshift over which damped systems can be detected  at a $> 5\sigma$ level 
(see Lanzetta \etal\ 1991 or Lanzetta \etal\ 1995). 
Figure~\ref{f_gz} shows the sensitivity function of the APM survey 
compared with three previous damped \lya surveys (Wolfe \etal\ 1986; 
Lanzetta \etal\ 1991; Lanzetta \etal\ 1995). 
The APM survey adds substantial redshift path 
for damped \lya absorption system surveys,
more than trebling the path surveyed for $z>3$.
The redshift path over which 
damped systems could be detected is crucial in estimating 
the cosmological mass density in neutral gas from the damped
systems (\cite{SMI95}).
Gaussians were fit to the lines selected by the algorithm,
and \nhi\ was estimated for features with W $>$20\AA.
Of the 32 measured, 11 have estimated
\nhi\ $\ge2\times10^{20}$ cm$^{-2}$ covering $2.8\le z \le 4.4$.
Only one candidate has estimated \nhi\ $\ge10^{21}$ cm$^{-2}$.
They are all listed in table 7
along with the QSOs with no candidates detected above this threshold.
The absorbers listed in table 7 are marked with a vertical slash 
in the spectra in figure~\ref{f_dla}. 
The damped \lya candidates 
with estimated column densities above the threshold of log \nhi\ $\ge$ 20.3 have
an asterisk after the column density in table 7 and a circle around the 
vertical slash in figure~\ref{f_dla}. 

To test this selection procedure, 
simulated high resolution spectra with damped systems
included at known column densities were degraded   
to 6\AA\ resolution and the column densities estimated with 
the above technique.  
The estimates were within $\pm 0.2 \times 10^{20}$ atoms cm$^{-2}$ 
of the real value. Two simulated QSOs 
are shown in figure~\ref{f_sim}. 
Panels (a) and (b) show a $z=3.86$, log \nhi\ $=$ 20.69 
damped \lya absorption system in a $z=4.37$ QSO at 1.6\AA\ 
and 6\AA\ resolution,  
respectively, with a signal-to-noise ratio of 25.  
Panels (c) and (d) show a $z=3.73$, log \nhi\ $=$ 20.15 
\lya absorption system in a $z=4.51$ QSO at 1.6\AA\ and 6\AA\ resolution,
respectively, with a signal-to-noise ratio of 10.

Many of the damped candidates have estimated \nhi\ near the 
statistical sample threshold of \nhi\ $= 2 \times 10^{20}$ atoms cm$^{-2}$
(\cite{Wolfe86}). 
Some of those with \nhi\ $< 2 \times 10^{20}$ will be confirmed as damped
and some of these with \nhi\ $\ge 2 \times 10^{20}$ will turn out to be
blends of weaker lines.  Higher resolution spectra
are essential to verify the individual column densities
though the estimates should give an accurate representation
of their distribution at high redshift.  
Three of the candidates have been observed at ESO 
with the NTT and these are discussed in Storrie-Lombardi, 
Irwin \& McMahon (1996). 
Preliminary results from 1.5\AA\ resolution spectra taken with
LRIS on the Keck telescope indicate that the rest of the 
candidates selected are representative
of the true distribution of HI column densities at high redshift 
(Storrie-Lombardi \& Wolfe, in preparation). 
When the follow-up spectroscopy is complete, 
this will result in a complete sample of damped absorbers
for z$>$3.5, increasing the confirmed numbers 
of these absorbers by  $\sim$ 20 percent and covering
an epoch crucial to understanding the formation of galaxies.

\section{Conclusions}
Intermediate resolution (5\AA) spectrophotometry are presented
for 31 QSOs with redshifts 3.9 $\le$ z $\le$ 4.7,  
28 from the APM Color Survey 
and 3 radio-selected objects. 
The spectra were surveyed to create new data sets
of intervening absorption lines systems.  
The QSOs display a wide variety of emission and absorption
line characteristics, with 5 exhibiting broad absorption lines 
and one with extremely strong emission lines (BR2248$-$1242).

This high redshift data set more than triples the $z>3$ redshift path
available for damped \lya absorption system surveys. 
Eleven candidate damped systems
have been identified covering the redshift range 
$2.8\le z \le 4.4$ (8 with $z>3.5$).
The redshift evolution, column density 
distribution function, and contribution to the cosmological
mass density from these systems is discussed in 
other papers (Storrie-Lombardi, Irwin, \& McMahon 1996; 
Storrie-Lombardi, McMahon, \& Irwin 1996).

The Lyman-limit systems in the QSOs with z $\ge$ 4.2 
are catalogued and the spectra presented. Their redshift
evolution has been discussed in a previous
paper (\cite{SMIH94}).  In addition, 
line lists for metal absorption line systems (\eg\ \civ\ and 
\mgii) are presented.
An analysis of the measured redshifts of the high ionization
emission lines with the low ionization lines shows them to be
blueshifted by $430\pm60$ km s$^{-1}$.  

\acknowledgments{We thank an anonymous referee and Mike Fall for suggestions
that improved the paper and Art Wolfe for the use of his
code in the automated damped candidate selection.  
We thank the PATT for time awarded to do
the observations with the William Herschel Telescope that made
this work possible.
LSL acknowledges support from an Isaac Newton Studentship, the Cambridge
Overseas Trust, and a University of California President's Postdoctoral
Fellowship.  RGM acknowledges the support of the Royal Society.
}


\clearpage

\figcaption[f_sky.ps]
{The 5\AA\ resolution flux calibrated sky spectrum from
QSO BRI1335$-0417$ taken with the red and blue arms of the
ISIS spectrograph at the WHT.
\label{f_sky}
}

\figcaption[f_bstar2.ps]
{B-star HR4468:
[B9.5, m$_{\rm V}=4.7$, exposure=1 second, 92 Apr 24, airmass=1.48]
\label{f_bstar2}
}

\figcaption[f_bstar.ps]
{B-star HD13679:
[B8,  m$_{\rm V}=6.8$, exposure=1 second, 93 Aug 21, airmass=1.056]
(a) wavelength calibrated counts spectrum, (b)
with atmospheric absorption features removed, and
(c) the spectrum in `a' divided by the spectrum in `b'.
The absorption spectrum in `c' is then divided into the QSO spectra
taken at a similar airmass to remove the atmospheric absorption features.
\label{f_bstar}
}

\figcaption[f_data1.ps]
{The final flux calibrated spectra. The z $>$ 4.2
QSOs used in the Lyman limit system evolution analysis
show the region blueward of 5500\AA\ magnified in the upper
left hand corner. All except BR0351$-$1034, BR0401$-$1711 and 
BR2237$-$0607 show the flux corrected for slit losses.
\label{f_data}
}

{Fig.~\ref{f_data} {\it continued.---}
The flux for BR0351$-$1034 has not been corrected for slit losses.} 

{Fig.~\ref{f_data} {\it continued.---}
The flux for BR0401$-$1711 has not been corrected for slit losses.} 

{Fig.~\ref{f_data} {\it continued.---}}

{Fig.~\ref{f_data} {\it continued.---}}

{Fig.~\ref{f_data} {\it continued.---}}

{Fig.~\ref{f_data} {\it continued.---}}

{Fig.~\ref{f_data} {\it continued.---}}

{Fig.~\ref{f_data} {\it continued.---}}

{Fig.~\ref{f_data} {\it continued.---}
The flux for BR2237$-$0607 has not been corrected for slit losses.} 

{Fig.~\ref{f_data} {\it continued.---}
The upper panel shows the entire spectrum of BR2248-1242.  The
lower panel has the \lya and \civ\ emission lines cut
off to show the additional lines visible in the spectrum.}

\figcaption[f_vdiff.ps]
{Histograms of the velocity difference relative to \civ\ for all
the measured emission lines are shown.  These are tabulated
in table 4.
Some of the very large differences of several
thousand kilometers per second are due to the difficulty in
accurately measuring some of the heavily absorbed emission lines,
\eg\ Ly$\alpha$.
\label{f_vdiff}
}

\figcaption[f_gz.ps]
{The sensitivity function, $g(z)$, of the
damped \lya absorber surveys.  This gives
the number of lines of sight along which a
damped system at redshift $z$ could be detected.
The APM survey adds substantial redshift path
for z$>$3.
\label{f_gz}
}

\figcaption[f_dla.ps]
{The \lya absorbers listed in table 7 are marked with a vertical slash 
in the spectra in this figure. It shows the \lya forest region on an expanded scale
for the QSOs shown in figure 4 in which an absorber was
measured.  The damped \lya candidates 
with estimated column densities above the threshold of log \nhi\ $\ge$ 20.3 have
an asterisk after the column density in table 7 and a circle around the 
vertical slash in the figure. 
\label{f_dla}
}

\figcaption[f_sim.ps]
{Two simulated QSOs with absorbers are shown. 
Panels (a) and (b) show a $z=3.86$, log \nhi\ $=$ 20.69 
damped \lya absorption system in a $z=4.37$ QSO at 1.6\AA\ 
and 6\AA\ resolution,
respectively, with a signal-to-noise ratio of 25.
Panels (c) and (d) show a $z=3.73$, log \nhi\ $=$ 20.15 
\lya absorption system in a $z=4.51$ QSO at 1.6\AA\ and 6\AA\ resolution,
respectively, with a signal-to-noise ratio of 10.
\label{f_sim}
}


\clearpage

\begin{thebibliography}{}

\bibitem[Bajtlik, Duncan, \& Ostriker \etal\ 1988]{BDO88}
Bajtlik, S., Duncan, R.C., \& Ostriker, J.P. 1988, \apj, 327, 570
\bibitem[Becker, Helfand, \& White 1992]{Becker92}
Becker, R.H., Helfand, D.J., \& White, R.L. 1992, \apj, 104, 531
\bibitem[Carswell \etal\ 1991]{RFC1331}
Carswell, R.F., \etal\ 1991, \apj, 381, L5
\bibitem[Carswell \etal\ 1987] {Carswell87}
Carswell, R.F., Webb, J.K., Baldwin, J.A., \& Atwood, B. 1987, \apj, 
  319, 709
\bibitem[Espey \etal\ 1989]{Espey89}
Espey, B.R., Carswell, R.F., Bailey, J.A., Smith, M.G. \&
Ward, M.J., 1989, \apj, 342, 666
\bibitem[Fall, Pei, \& McMahon 1989]{FPM89}
Fall, S.M., Pei, Y.C., \& McMahon, R.G. 1989, \apj, 341, L5
\bibitem[Filippenko 1982]{Filippenko82}
Filippenko, A.V. 1982, \pasp, 94, 715
\bibitem[Hayes \& Latham 1975]{HL75}
Hayes, D.S. \& Latham, D.W., 1975, \apj, 197, 593
\bibitem[Hirshfeld, Sinnott, \& Ochsenbien 1991]{Skycat}
Hirshfeld, A., Sinnott, R.W., \& Ochsenbien, F. 1991, Sky Catalogue
  2000.0 Cambridge University Press
\bibitem[Hoffleit \& Jaschek 1982]{BrightStarCat}
Hoffleit, D. \& Jaschek, C. 1982, Bright Star Catalogue Yale University
  Observatory
\bibitem[Hook \etal\ 1995]{Hook95} Hook, I.M.,
McMahon, R.G., Patnaik, A.R., Browne, W.A., Wilkinson, P.N.,
Irwin, M.J., \& Hazard, C., 1995, \mnras, 273, L63
\bibitem[Hunstead \etal\ 1986]{Hunstead86}
Hunstead, R.W., Murdoch, H.S., Peterson, B.A., Blades, J.C., Jauncey,
  D.L., Wright, A.E., Pettini, M., \& Savage, A. 1986, \apj, 305, 496
\bibitem[Irwin, McMahon, \& Hazard 1991] {IMH91}
Irwin, M.J., McMahon, R.G., \& Hazard, C. 1991, in Astronomical Society
  of the Pacific Conference Series, Vol. 21, ed. D.~Crampton (San Francisco:
  Astronomical Society of the Pacific), 117
\bibitem[Lanzetta 1991]{Lanzetta91a}
Lanzetta, K.M. 1991, \apj, 375, 1
\bibitem[Lanzetta, Wolfe, \& Turnshek 1995]{LWT95}
Lanzetta, K.M., Wolfe, A.M., \& Turnshek, D.A. 1995, \apj, 440, 435
\bibitem[Lanzetta \etal\ 1991]{LWTLMH91}
Lanzetta, K.M., Wolfe, A.M., Turnshek, D.A., Lu, L., McMahon, R.G.,
  \& Hazard, C. 1991, ApJS, 77, 1 
\bibitem[McMahon, Irwin, \& Hazard 1992]{Lens0952}
McMahon, R.G., Irwin, M.J., \& Hazard, C. 1992, Gemini Issue 36, 1
\bibitem[McMahon \etal\ 1994]{RGMiram94} McMahon, R.G.,
Omont, A., Bergeron, J., Kreysa, E., \& Haslam, C.G.T.,
1994, \mnras, 267, L9
\bibitem[Oke 1969]{Oke69} Oke, J.B. 1969, \pasp, 81, 11
\bibitem[Oke 1974]{Oke74} Oke, J.B. 1974, \apjs, 27, 21
\bibitem[Oke \& Gunn 1983]{OkeGunn83}
Oke, J.B. \& Gunn, J.E. 1983, \apj, 266, 713
\bibitem[Pettini, Boksenberg, \& Hunstead 1990]{PBH90}
Pettini, M., Boksenberg, A., \& Hunstead, R.W. 1990, \apj, 348, 48
\bibitem[Rauch \etal\ 1990] {Rauch90}
Rauch, M., Carswell, R.F., Robertson, J.G., Shaver, P.A., \& Webb,
  J.K. 1990, \mnras, 242, 698
\bibitem[Sargent, Steidel, \& Boksenberg 1989]{SSB}
Sargent, W.L.W., Steidel, C.C., \& Boksenberg, A. 1989, \apjs, 79, 703
\bibitem[Steidel \& Sargent 1991]{SS91}
Steidel, C.C. \& Sargent, W.L.W. 1991, \apj, 382, 433
\bibitem[Stengler-Larrea \etal\ 1995]{Stengler95}
Stengler-Larrea, E.A., Boksenberg, A., Steidel, C.C.,
Sargent, W.L.W., Bahcall, J.N., Bergeron, J., Hartig, G.F.,
Januzzi, B.T., Kirhakos, S., Savage, B.D., Schneider, D.P.,
Turnshek, D.A., \& Weymann, R.J., 1995, \apj, 444, 64
\bibitem[Storrie-Lombardi, Irwin, \& McMahon 1996]{SIM95}
Storrie-Lombardi, L.J., Irwin, M.J., \& McMahon, R.G. 1996, MNRAS, submitted
\bibitem[Storrie-Lombardi, McMahon, \& Irwin 1996]{SMI95}
Storrie-Lombardi, L.J., McMahon, R.G., \& Irwin, M.J. 1996, ApJ, submitted
\bibitem[Storrie-Lombardi \etal\ 1994]{SMIH94}
Storrie-Lombardi, L.J., McMahon, R.G., Irwin, M.J., \& Hazard, C. 1994,
  \apj, 427, L13
\bibitem[Turnshek \etal\ 1989]{TWLBetal89}
Turnshek, D.A., Wolfe, A.M., Lanzetta, K.M., Briggs, F.H., Cohen,
  R.D., Foltz, C.B., Smith, H.E., \& Wilkes, B.J. 1989, \apj, 344, 567
\bibitem[Tytler \& Fan 1992]{TF92}
Tytler, D. \& Fan, X.M. 1992, \apjs, 79, 1
\bibitem[Williger \etal\ 1994]{Williger94}
Williger, G.M., Baldwin, J.A., Carswell, R.F., Cooke, A.J., Hazard,
  C., Irwin, M.J., McMahon, R.G., \& Storrie-Lombardi, L.J. 1994, \apj,
  428, 574
\bibitem[Wampler \etal\ 1996]{Wampler96}
Wampler, E.J., Williger, G.M., Baldwin, J.A., Carswell, R.F.,
Hazard, C., \& McMahon, R.G., 1996, A\&A, submitted
\bibitem[White \& Becker 1992]{WB92}
White, R.L. \& Becker, R.H., 1992, \apjs, 79, 331
\bibitem[Wolfe 1987]{Wolfe87}
Wolfe, A.M. 1987, Proc.~Phil.~Trans.~Roy.~Soc., 320, 503
\bibitem[Wolfe \etal\ 1995]{WLFC95}
Wolfe, A.M., Lanzetta, K.M., Foltz C.B., \& Chaffee F.H., 
1995, ApJ, 454, 698
\bibitem[Wolfe \etal\ 1986]{Wolfe86}
Wolfe, A.M., Turnshek, D.A., Smith, H.E., \& Cohen, R.D. 1986, \apjs, 61, 249 
\end{thebibliography}
\end{document}